\documentclass[preprintnumbers,prd,twocolumn,showpacs,floatfix,preprintnumbers,letterpaper,amsmath,amssymb,nofootinbib,superscriptaddress]{revtex4}
\usepackage{graphicx}
\usepackage{epsfig}
\usepackage{bm}
\usepackage{amsfonts}

\newcommand{\la}{\lambda}

\newcommand{\bear}{\begin{eqnarray}}

\newcommand{\eear}{\end{eqnarray}}

\newbox\pippobox

\def\6{\partial}

\def\a{\alpha}

\def\sq
\def\a{\alpha}

\def\dn{\Delta N_{\nu}}

\newcommand {\lla} {\ {\raise-.5ex\hbox{$\buildrel<\over\sim$}}\ }

\def\be{\begin{equation}}
\def\ee{\end{equation}}
\def\ba{\begin{eqnarray}}
\def\ea{\end{eqnarray}}
\def\w{\omega}
\renewcommand{\(}{\left(}
\renewcommand{\)}{\right)}
\renewcommand{\[}{\left[}
\renewcommand{\]}{\right]}

\begin{document}

\title{Overall observational constraints on the running parameter $\lambda$ of Ho\v{r}ava-Lifshitz gravity}

\author{Sourish Dutta}
\email{sourish.d@gmail.com} \affiliation{Department of Physics and
Astronomy, Vanderbilt University, Nashville, TN  ~~37235}

\author{Emmanuel N. Saridakis}
 \email{msaridak@phys.uoa.gr}
 \affiliation{College of Mathematics and Physics,\\ Chongqing University of Posts and
Telecommunications, Chongqing, 400065, P.R. China }

\begin{abstract}
We use observational data from  Type Ia Supernovae (SNIa), Baryon
Acoustic Oscillations (BAO), and Cosmic Microwave Background
(CMB), along with requirements of Big Bang Nucleosynthesis (BBN),
to constrain the running parameter $\lambda$ of
Ho\v{r}ava-Lifshitz gravity, which determines the flow between the
Ultra-Violet and the Infra-Red. We consider both the detailed and
non-detailed balance versions of the gravitational sector, and we
include the matter and radiation sectors. Allowing for variation
of all the parameters of the theory, we construct the likelihood
contours and we conclude that in $1\sigma$ confidence $\lambda$ is
restricted to $|\lambda-1|\lesssim0.02$, while its best fit value
is $|\lambda_{b.f}-1|\approx0.002$.
 Although this observational analysis restricts the running parameter $\lambda$ very close
to its IR value 1, it does not enlighten the discussion about the
theory's possible conceptual and theoretical problems.
\end{abstract}


\maketitle

\section{Introduction}

Ho\v{r}ava recently proposed a power-counting renormalizable,
Ultra-Violet (UV) complete theory of gravity
\cite{hor2,hor1,hor3,hor4}. Although presenting an Infra-Red (IR)
fixed point, namely General Relativity, in the  UV the theory
possesses a fixed point with an anisotropic, Lifshitz scaling
between time and space. Since then there has been a significant
progress in examining  the properties of the theory itself,
including various extensions of the original basic version
\cite{Volovik:2009av,Cai:2009ar,Cai:2009dx,Orlando:2009en,Nishioka:2009iq,Konoplya:2009ig,Charmousis:2009tc,Li:2009bg,Nojiri:2009th,Kluson:2009rk,Visser:2009fg,Sotiriou:2009gy,
Sotiriou:2009bx,Germani:2009yt,Chen:2009bu,Chen:2009ka,Shu:2009gc,Bogdanos:2009uj,Afshordi:2009tt,Myung:2009ur,Alexandre:2009sy,
Blas:2009qj,Capasso:2009fh,Chen:2009vu,Kluson:2009xx,Kluson:2010aw}.
Additionally, application of Ho\v{r}ava-Lifshitz gravity as a
cosmological framework gives rise to Ho\v{r}ava-Lifshitz cosmology
\cite{Calcagni:2009ar,Kiritsis:2009sh}, and in particular one can
study specific solution subclasses
\cite{Lu:2009em,Nastase:2009nk,Colgain:2009fe,Ghodsi:2009rv,Minamitsuji:2009ii,Ghodsi:2009zi,Wu:2009ah,Cho:2009fc,Boehmer:2009yz,Momeni:2009au,Setare:2009sw,Kiritsis:2009vz,Cai:2010ud,Chaichian:2010yi,Ali:2010sv},
the phase-space behavior
\cite{Carloni:2009jc,Leon:2009rc,Myung:2009if,Bakas:2009ku,Myung:2010qg},
the gravitational wave production
\cite{Mukohyama:2009zs,Takahashi:2009wc,Koh:2009cy,Park:2009gf,Park:2009hg,Myung:2009ug},
the perturbation spectrum
\cite{Mukohyama:2009gg,Piao:2009ax,Gao:2009bx,Chen:2009jr,Gao:2009ht,Cai:2009hc,Wang:2009yz,Kobayashi:2009hh,Wang:2009azb,Kobayashi:2010eh},
the matter bounce
\cite{Brandenberger:2009yt,Brandenberger:2009ic,Cai:2009in,Suyama:2009vy,Czuchry:2009hz,Gao:2009wn},
the black hole properties
\cite{Danielsson:2009gi,Cai:2009pe,Myung:2009dc,Kehagias:2009is,Mann:2009yx,Bertoldi:2009vn,Park:2009zra,Castillo:2009ci,BottaCantcheff:2009mp,Lee:2009rm,Varghese:2009xm,Kiritsis:2009rx,Majhi:2009xh,Greenwald:2009kp,Lobo:2010hv},
the dark energy phenomenology
\cite{Saridakis:2009bv,Park:2009zr,Appignani:2009dy,Setare:2009vm,Garattini:2009ns,Jamil:2010vr},
the astrophysical phenomenology
\cite{Kim:2009dq,Iorio:2009qx,Iorio:2009ek,Izumi:2009ry,Harko:2009qr},
the thermodynamic properties
\cite{Wang:2009rw,Cai:2009qs,Cai:2009ph} etc. However, despite
this extended research, there are still many ambiguities if
Ho\v{r}ava-Lifshitz gravity is reliable and capable of a
successful description of the gravitational background of our
world, as well as of the cosmological behavior of the universe
\cite{Charmousis:2009tc,Li:2009bg,Sotiriou:2009bx,Bogdanos:2009uj,Blas:2009yd,Koyama:2009hc,Papazoglou:2009fj,Henneaux:2009zb,Gong:2010xp}.

Although the discussion about the foundations and the possible
conceptual and phenomenological problems of Ho\v{r}ava-Lifshitz
gravity and cosmology is still open in the literature, it is
worthy to examine the possible constraints that observational and
cosmological data could impose on the parameters of the scenario
(of the basic as well as of the various extended versions). In
such an investigation a reasonable assumption is to set the
running parameter $\lambda$ to its IR value 1, since all
observations lie deep inside the IR. Thus, under this assumption
in \cite{Kim:2009dq,Iorio:2009qx,Iorio:2009ek} the authors used
Solar System observations in order to constrain some of the
remaining model parameters of the basic version of
Ho\v{r}ava-Lifshitz cosmology under detailed balance, while in
\cite{Harko:2009qr}  the analysis was performed including a soft
detailed-balance breaking. Similarly, in \cite{Dutta:2009jn} we
used  cosmological observations (Type Ia Supernovae (SNIa), Baryon
Acoustic Oscillations (BAO) and Cosmic Microwave Background (CMB)
ones, together with Big Bang Nucleosynthesis conditions) in order
to impose complete constraints on all the parameters of the basic
version of Ho\v{r}ava-Lifshitz cosmology and construct the
corresponding contour plots, with or without the detailed-balance
condition.

Although setting  $\lambda$ to its IR value is a first and
reasonable assumption, one could go beyond it, and examine the
observational constraints that the data could impose on $\lambda$
itself. However, allowing $\lambda$ varying, that is preserving
the Lorentz invariance breaking (which is restored in the exact IR
value $\lambda=1$), one must take into account that in theories
with Lorentz invariance breaking the ``gravitational'' Newton's
constant $G_{\rm grav}$, that is the one that is present in the
gravitational action, does not coincide with the ``cosmological''
Newton's constant $G_{\rm cosmo}$, that is the one that is present in
Friedmann equations \cite{Carroll:2004ai}. Thus, in the case of
Ho\v{r}ava-Lifshitz gravity one could use the deviation between
$G_{\rm grav}$ and $G_{\rm cosmo}$, as it is constrained by measurements
of the primordial abundance of He$^4$ \cite{Carroll:2004ai}, in
order to extract an upper bound on $|\lambda-1|$. This approach
was followed in \cite{Blas:2009qj,Papazoglou:2009fj} with the
result $0<|\lambda-1|\lesssim0.1$. However, it is obvious that
such an approach can only provide a crude upper bound on
$|\lambda-1|$, since it considers that all the other parameters of
the theory remain constant. The correct approach should be to
perform a systematic investigation, allowing for simultaneous
variations of all model parameters, and constrain all of them
using observations.

In the present work we are interested in performing such an
holistic observational constraining of all the parameters of the
basic version of Ho\v{r}ava-Lifshitz cosmology, and especially of
the running parameter $\lambda$, using  SNIa, BAO and CMB
cosmological observations  in order to construct the corresponding
probability contour-plots. Furthermore, in order to be general and
model-independent, we perform our analysis with and without the
detailed-balance condition. The plan of the work is the following:
In section \ref{model} we present the basic ingredients of
Ho\v{r}ava-Lifshitz cosmology, extracting the Friedmann equations,
and describing the dark matter and dark energy dynamics. In
section \ref{Observational constraints} we constrain both the
detailed-balance  and the beyond-detailed-balance formulations
using cosmological observations, and we present the corresponding
likelihood contours. Finally, in section \ref{conclusions} we
summarize the obtained results.

\section{Ho\v{r}ava-Lifshitz cosmology}
\label{model}

In this section we briefly review the scenario where the
cosmological evolution is governed by Ho\v{r}ava-Lifshitz gravity
\cite{Calcagni:2009ar,Kiritsis:2009sh}. The dynamical variables
are the lapse and shift functions, $N$ and $N_i$ respectively, and
the spatial metric $g_{ij}$ (roman letters indicate spatial
indices). In terms of these fields the full metric is written as:
\begin{eqnarray}
ds^2 = - N^2 dt^2 + g_{ij} (dx^i + N^i dt ) ( dx^j + N^j dt ) ,
\end{eqnarray}
where indices are raised and lowered using $g_{ij}$. The scaling
transformation of the coordinates reads:
$
 t \rightarrow l^3 t~~~{\rm and}\ \ x^i \rightarrow l x^i
$.

\subsection{Detailed Balance}

The gravitational action is decomposed into a kinetic and a
potential part as $S_g = \int dt d^3x \sqrt{g} N ({\cal L}_K+{\cal
L}_V)$. The assumption of detailed balance \cite{hor3}
  reduces the possible terms in the Lagrangian, and it allows
for a quantum inheritance principle \cite{hor2}, since the
$(D+1)$-dimensional theory acquires the renormalization properties
of the $D$-dimensional one. Under the detailed balance condition
 the full action of Ho\v{r}ava-Lifshitz gravity is given by
\begin{eqnarray}
 S_g &=&  \int dt d^3x \sqrt{g} N \left\{
\frac{2}{\kappa^2}
(K_{ij}K^{ij} - \lambda K^2) \ \ \ \ \ \ \ \ \ \ \ \ \ \ \ \ \  \right. \nonumber \\
&+&\left.\frac{\kappa^2}{2 w^4} C_{ij}C^{ij}
 -\frac{\kappa^2 \mu}{2 w^2}
\frac{\epsilon^{ijk}}{\sqrt{g}} R_{il} \nabla_j R^l_k +
\frac{\kappa^2 \mu^2}{8} R_{ij} R^{ij}
     \right. \nonumber \\
&+&\left.    \frac{\kappa^2 \mu^2}{8( 3 \lambda-1)} \left[ \frac{1
- 4 \lambda}{4} R^2 + \Lambda  R - 3 \Lambda ^2 \right] \right\},
\label{acct}
\end{eqnarray}
where
\begin{eqnarray}
K_{ij} = \frac{1}{2N} \left( {\dot{g_{ij}}} - \nabla_i N_j -
\nabla_j N_i \right)
\end{eqnarray}
is the extrinsic curvature and
\begin{eqnarray} C^{ij} \, = \, \frac{\epsilon^{ijk}}{\sqrt{g}} \nabla_k
\bigl( R^j_i - \frac{1}{4} R \delta^j_i \bigr)
\end{eqnarray}
the Cotton tensor, and the covariant derivatives are defined with
respect to the spatial metric $g_{ij}$. $\epsilon^{ijk}$ is the
totally antisymmetric unit tensor, $\lambda$ is a dimensionless
constant and the variables $\kappa$, $w$ and $\mu$ are constants
with mass dimensions $-1$, $0$ and $1$, respectively. Finally, we
mention that in action (\ref{acct}) we have already performed the
usual analytic continuation of the parameters $\mu$ and $w$ of the
original version of Ho\v{r}ava-Lifshitz gravity, since such a
procedure is required in order to obtain a realistic cosmology
\cite{Lu:2009em,Minamitsuji:2009ii,Wang:2009rw,Park:2009zra}
(although it could fatally affect the gravitational theory
itself). Therefore, in the present work $\Lambda $ is a positive
constant, which as usual is related to the cosmological constant
in the IR limit.

In order to add the matter component (including both dark and
baryonic matter)  in the theory one can follow two equivalent
approaches. The first is to introduce a scalar field
\cite{Calcagni:2009ar,Kiritsis:2009sh} and thus attribute to dark
matter a dynamical behavior, with its energy density $\rho_m$ and
pressure $p_m$ defined through the field kinetic and potential
energy. Although such an approach is theoretically robust, it is
not suitable from the phenomenological point of view since it
requires specially-designed matter-potentials in order to acquire
an almost constant matter equation-of-state parameter
($w_m=p_m/\rho_m$) as it is suggested by observations. In the
second approach one adds a cosmological stress-energy tensor to
the gravitational field equations, by demanding to recover the
usual general relativity formulation in the low-energy limit
\cite{Sotiriou:2009bx,Chaichian:2010yi,Carloni:2009jc}. Thus, this
matter-tensor is a hydrodynamical approximation with $\rho_m$ and
$p_m$ (or $\rho_m$ and $w_m$) as parameters. Similarly, one can
additionally include the standard-model-radiation component
(corresponding to photons and neutrinos), with the additional
parameters $\rho_r$ and $p_r$ (or $\rho_r$ and $w_r$). Such an
approach, although not fundamental, is better for a
phenomenological analysis, such the one performed in this work.

In order to investigate cosmological frameworks, we impose the
projectability condition \cite{Charmousis:2009tc} and we use an
FRW metric
\begin{eqnarray}
N=1~,~~g_{ij}=a^2(t)\gamma_{ij}~,~~N^i=0~,
\end{eqnarray}
with
\begin{eqnarray}
\gamma_{ij}dx^idx^j=\frac{dr^2}{1- K r^2}+r^2d\Omega_2^2~,
\end{eqnarray}
where $ K<,=,> 0$ corresponding  to open, flat, and closed
universe respectively (we have adopted the convention of taking
the scale factor $a(t)$ to be dimensionless and the curvature
constant $ K$ to have mass dimension 2). By varying $N$ and
$g_{ij}$, we extract the Friedmann equations:
\begin{eqnarray}\label{Fr1fluid}
H^2 &=&
\frac{\kappa^2}{6(3\la-1)}\Big(\rho_m+\rho_r\Big)+\nonumber\\
&+&\frac{\kappa^2}{6(3\la-1)}\left[ \frac{3\kappa^2\mu^2
K^2}{8(3\lambda-1)a^4} +\frac{3\kappa^2\mu^2\Lambda
^2}{8(3\lambda-1)}
 \right]-\nonumber\\
 &-&\frac{\kappa^4\mu^2\Lambda  K}{8(3\lambda-1)^2a^2} \ ,
\end{eqnarray}
\begin{eqnarray}\label{Fr2fluid}
\dot{H}+\frac{3}{2}H^2 &=&
-\frac{\kappa^2}{4(3\la-1)}\Big(w_m\rho_m+w_r\rho_r\Big)-\nonumber\\
&-&\frac{\kappa^2}{4(3\la-1)}\left[\frac{\kappa^2\mu^2
K^2}{8(3\lambda-1)a^4} -\frac{3\kappa^2\mu^2\Lambda
^2}{8(3\lambda-1)}
 \right]-\nonumber\\
 &-&\frac{\kappa^4\mu^2\Lambda  K}{16(3\lambda-1)^2a^2}\ ,
\end{eqnarray}
where  $H\equiv\frac{\dot a}{a}$ is the Hubble parameter. As
usual, $\rho_m$ (dark plus baryonic matter) follows the standard
evolution equation
\begin{eqnarray}\label{rhodotfluid}
&&\dot{\rho}_m+3H(\rho_m+p_m)=0,
\end{eqnarray}
while $\rho_r$ (standard-model radiation) follows
\begin{eqnarray}\label{rhodotfluidrad}
&&\dot{\rho}_r+3H(\rho_r+p_r)=0.
\end{eqnarray}

Lastly, concerning the dark-energy sector we can define
\begin{equation}\label{rhoDE}
\rho_{DE}\equiv \frac{3\kappa^2\mu^2 K^2}{8(3\lambda-1)a^4}
+\frac{3\kappa^2\mu^2\Lambda ^2}{8(3\lambda-1)}
\end{equation}
\begin{equation}
\label{pDE} p_{DE}\equiv \frac{\kappa^2\mu^2
K^2}{8(3\lambda-1)a^4} -\frac{3\kappa^2\mu^2\Lambda
^2}{8(3\lambda-1)}.
\end{equation}
The term proportional to $a^{-4}$ is the usual ``dark radiation
term'', present in Ho\v{r}ava-Lifshitz cosmology
\cite{Calcagni:2009ar,Kiritsis:2009sh}, while the constant term is
just the explicit cosmological constant. Therefore, in expressions
(\ref{rhoDE}),(\ref{pDE}) we have defined the energy density and
pressure for the effective dark energy, which incorporates the
aforementioned contributions. Finally, note that using
(\ref{rhoDE}),(\ref{pDE}) it is straightforward to show that these
 dark energy quantities satisfy the
standard evolution equation:
\begin{eqnarray}
\label{DEevol} &&\dot{\rho}_{DE}+3H(\rho_{DE}+p_{DE})=0.
\end{eqnarray}

Using the above definitions, we can re-write the Friedmann
equations (\ref{Fr1fluid}),(\ref{Fr2fluid}) in the standard form:
\begin{equation}
\label{Fr1b} H^2 =
\frac{\kappa^2}{6(3\la-1)}\Big[\rho_m+\rho_r+\rho_{DE}\Big]-
\frac{\kappa^4\mu^2\Lambda  K}{8(3\lambda-1)^2a^2}
\end{equation}
\begin{equation}
\label{Fr2b} \dot{H}+\frac{3}{2}H^2 =
-\frac{\kappa^2}{4(3\la-1)}\Big[p_m+p_r+p_{DE}
 \Big]-\frac{\kappa^4\mu^2\Lambda  K}{16(3\lambda-1)^2a^2}.
\end{equation}
Therefore, if we require these expressions to coincide with the
standard Friedmann equations, in units where $c=1$  we set
\cite{Calcagni:2009ar,Kiritsis:2009sh}:
\begin{eqnarray}
G_{\rm cosmo}&=&\frac{\kappa^2}{16\pi(3\lambda-1)}\nonumber\\
\frac{\kappa^4\mu^2\Lambda}{8(3\lambda-1)^2}&=&1,
\label{simpleconstants0}
\end{eqnarray}
where $G_{\rm cosmo}$ is the ``cosmological'' Newton's constant. Note
that as we said in the Introduction, in theories with Lorentz
invariance breaking $G_{\rm cosmo}$ does not coincide with the
 ``gravitational'' Newton's constant
$G_{\rm grav}$, unless Lorentz invariance is restored
\cite{Carroll:2004ai}. For completeness we mention that in our
case
\begin{eqnarray}
G_{\rm grav}=\frac{\kappa^2}{32\pi}\label{Ggrav},
\end{eqnarray}
as it can be straightforwardly read from the action (\ref{acct})
(our definitions of $G_{\rm cosmo}$, $G_{\rm grav}$ coincide with those of
\cite{Blas:2009qj,Papazoglou:2009fj}). Thus, it becomes obvious
that in the IR ($\lambda=1$), where Lorentz invariance is
restored, $G_{\rm cosmo}$ and $G_{\rm grav}$ coincide.

\subsection{Beyond Detailed Balance}

The above formulation of Ho\v{r}ava-Lifshitz cosmology has been
performed under the imposition of the detailed-balance condition.
However, in the literature there is a discussion whether this
condition leads to reliable results or if it is able to reveal the
full information of Ho\v{r}ava-Lifshitz
 gravity \cite{Calcagni:2009ar,Kiritsis:2009sh}. Thus, one
 should study also the Friedmann equations in the case
 where detailed balance is relaxed. In such a case one can in
 general write
 \cite{Charmousis:2009tc,Sotiriou:2009bx,Bogdanos:2009uj,Carloni:2009jc,Leon:2009rc}:
\begin{eqnarray}\label{Fr1c}
H^2 &=&
\frac{2\sigma_0}{(3\la-1)}\Big(\rho_m+\rho_r\Big)+\nonumber\\
&+&\frac{2}{(3\la-1)}\left[ \frac{\sigma_1}{6}+\frac{\sigma_3
K^2}{6a^4} +\frac{\sigma_4 K}{6a^6}
 \right]+\nonumber\\&+&\frac{\sigma_2}{3(3\la-1)}\frac{ K}{a^2}
\end{eqnarray}
\begin{eqnarray}\label{Fr2c}
\dot{H}+\frac{3}{2}H^2 &=&
-\frac{3\sigma_0}{(3\la-1)}\Big(w_m\rho_m+w_r\rho_r\Big)-\nonumber\\
&-&\frac{3}{(3\la-1)}\left[ -\frac{\sigma_1}{6}+\frac{\sigma_3
K^2}{18a^4} +\frac{\sigma_4 K}{6a^6}
 \right]+\nonumber\\&+&
 \frac{\sigma_2}{6(3\la-1)}\frac{ K}{a^2},
\end{eqnarray}
where $\sigma_0\equiv \kappa^2/12$, and the constants $\sigma_i$
are arbitrary (with $\sigma_2$ being negative). Note that one
could absorb the factor of $6$ in redefined parameters, but we
prefer to keep it in order to coincide with the notation of
\cite{Sotiriou:2009bx,Carloni:2009jc}. As we observe, the effect
of the detailed-balance relaxation is the decoupling of the
coefficients, together with the appearance of a term proportional
to $a^{-6}$. In this case the corresponding quantities for dark
energy are generalized to
\begin{eqnarray}\label{rhoDEext}
&&\rho_{DE}|_{_\text{non-db}}\equiv
\frac{\sigma_1}{6}+\frac{\sigma_3 K^2}{6a^4} +\frac{\sigma_4
K}{6a^6}
\\
&&\label{pDEext} p_{DE}|_{_\text{non-db}}\equiv
-\frac{\sigma_1}{6}+\frac{\sigma_3 K^2}{18a^4} +\frac{\sigma_4
K}{6a^6}.
\end{eqnarray}
Again, it is easy to show that
\begin{eqnarray}\label{rhodotfluidnd}
\dot{\rho}_{DE}|_{_\text{non-db}}+3H(\rho_{DE}|_{_\text{non-db}}+p_{DE}|_{_\text{non-db}})=0.
\end{eqnarray}
 Finally, if we force (\ref{Fr1c}),(\ref{Fr2c}) to coincide with
 the standard Friedmann equations, we result to:
\begin{eqnarray}
&&G_{\rm cosmo}=\frac{6\sigma_0}{8\pi(3\lambda-1)}\nonumber\\
&&\sigma_2=-3(3\lambda-1), \label{simpleconstants0nd}
\end{eqnarray}
while in this case the ``gravitational'' Newton's constant
$G_{\rm grav}$ reads \cite{Sotiriou:2009bx}:
\begin{eqnarray}
G_{\rm grav}=\frac{6\sigma_0}{16\pi}\label{Ggravbdb}.
\end{eqnarray}

\section{Observational constraints}
\label{Observational constraints}

Having presented the cosmological equations of a universe governed
by Ho\v{r}ava-Lifshitz gravity, both with and without the
detailed-balance condition, we now proceed to study  the
observational constraints on the model parameters. This is
performed in the following two subsections, for the detailed and
non-detailed balance scenarios separately.  We mention that,
contrary to \cite{Dutta:2009jn}, in this work we allow the running
parameter $\lambda$ to vary too, and in order to be general enough
we do not use any theoretical argument to restrict it in any
specific interval, handling it as completely free.

\subsection{Constraints on Detailed-Balance scenario}

We work in the usual units suitable for observational comparisons,
namely setting  $8\pi G_{\rm grav}=1$ (we have already set $c=1$ in
order to obtain (\ref{simpleconstants0})). This allows us to
reduce the parameter space, since in this case (\ref{Ggrav}) gives
\begin{eqnarray}
\kappa^2=4
 \label{simpleconstants0help},
\end{eqnarray}
and thus (\ref{simpleconstants0}) lead to:
\begin{eqnarray}
G_{\rm cosmo}&=&\frac{1}{4\pi(3\lambda-1)}\nonumber\\
 \mu^2\Lambda&=&\frac{(3\lambda-1)^2}{2}. \label{simpleconstants}
\end{eqnarray}
 Inserting these relations into Friedmann equation
(\ref{Fr1fluid}) we obtain
\begin{equation}
\label{FriedmanDB}
 H^2=\frac{2}{3(3\lambda-1)}\Big(\rho_m+\rho_r\Big)+\frac13\left(\frac{3 K^2}{2\Lambda
a^4}+\frac{3\Lambda}{2}\right)-\frac{ K}{a^2}.
 \end{equation}
In terms of the usual density parameters
($\Omega_m\equiv\rho_m/(3H^2)$, $\Omega_ K\equiv - K/(H^2a^2)$,
$\Omega_r\equiv\rho_r/(3H^2)$) this expression becomes:
\begin{equation}
 1-\frac{2}{(3\lambda-1)}\Big(\Omega_m+\Omega_r\Big)-\Omega_{ K}=\frac{1}{
H^2}\left(\frac{ K^2}{2\Lambda a^4}+\frac{\Lambda}{2}\right).
 \end{equation}
Applying this relation at present time and setting the current
scale factor $a_0=1$ we obtain:
\begin{equation}
\label{Fr0detb}
 1-\frac{2}{(3\lambda-1)}\Big(\Omega_{m0}+\Omega_{r0}\Big)-\Omega_{ K0}=\frac{1}{
H_0^2}\left(\frac{ K^2}{2\Lambda }+\frac{\Lambda}{2}\right),
 \end{equation}
where a $0$-subscript denotes the present value of the
corresponding quantity. Note that $\Omega_{m0}$  includes
contributions from both baryons $\Omega_{b0}$ as well as dark
matter $\Omega_{\rm DM0}$.

In order to proceed to the elaboration of observational data,  we
consider as usual the matter (dark plus baryonic) component to be
dust, that is $w_m\approx0$, and similarly for the standard-model
radiation we consider $w_r=1/3$, where both assumptions are valid
in the epochs in which observations focus. Therefore, the
corresponding evolution equations
(\ref{rhodotfluid}),(\ref{rhodotfluidrad}) give
$\rho_m=\rho_{m0}/a^3$ and $\rho_r=\rho_{r0}/a^4$ respectively.
Finally, it proves convenient to use the redshift $z$ as the
independent variable instead of the scale factor ($1+z\equiv
a_0/a=1/a$). Inserting these into Friedmann equation
(\ref{FriedmanDB}) we obtain
\begin{eqnarray}
H^2&=&H_{0}^2\Big\{\frac{2}{(3\lambda-1)}\Big[\Omega_{m0}(1+z)^3+\Omega_{r0}(1+z)^4\Big]+\nonumber\\
&\ &+\Omega_{ K0}(1+z)^2+\Big[\omega+\frac{\Omega_{ K
0}^2}{4\omega}(1+z)^4\Big] \Big\},
 \label{Frdbfinal}
\end{eqnarray}
  where we have also introduced  the
dimensionless parameter
\begin{eqnarray}
\omega\equiv\frac{\Lambda}{2 H_0^2}.
 \end{eqnarray}
 Thus, the constraint
(\ref{Fr0detb}) can be rewritten as:
  \be
\label{cond1}
\frac{2}{(3\lambda-1)}\Big(\Omega_{m0}+\Omega_{r0}\Big)+\Omega_{K0}+\omega+\frac{\Omega_{K0}^2}{4\omega}=1.
 \ee

We remind that the term $\Omega_{ K 0}^2/(4\omega)$ is the
coefficient of the dark radiation term, which is a characteristic
feature of the Ho\v{r}ava-Lifshitz gravitational background. Since
this dark radiation component has been present also during the
time of nucleosynthesis, it is subject to bounds from Big Bang
Nucleosynthesis (BBN). As discussed in more details in the
Appendix of \cite{Dutta:2009jn}, if the upper limit on the total
amount of dark radiation allowed during BBN is expressed through
the parameter $\Delta N_\nu$ of the effective neutrino species
\cite{BBNrefs,BBNrefs1,BBNrefs2,Malaney:1993ah}, then we obtain
the following constraint :
  \be
\label{cond2}
 \frac{\Omega_{ K 0}^2}{4\omega}=0.135\dn \Omega_{r0}.
  \ee
In this work, in order to ensure consistency with BBN, we adopt an upper limit of $\dn\leq2.0$ following \cite{BBNrefs2}.

\begin{center}
\begin{table}[ht]
    \centering
    \scalebox{.95}{
        \begin{tabular}{|c|c|c|c|c|c|c|}
        \hline \textbf{$K$} &
         $\kappa^2/(8\pi G_{\rm grav})$ & \textbf{$\left(1/H_0^2\right)\Lambda $}& $ \left(8\pi G_{\rm grav} H_0\right)\mu$ & $\lambda$ & $\dn$\\\hline
        $>$0&     4      & $(0,\,1.46 )$ & $(1.37,\infty)$ & (0.98, 1.01) & (0, 0.32)\\\hline
           $<$0 &   4      & $(0,\,1.46) $ & $(1.8,\infty)$ & (0.97, 1.01)& (0, 0.68) \\\hline
        \end{tabular}}
    \caption{1$\sigma$ limits on the parameter values for the detailed-balance scenario, for positive and negative curvature.The cosmological parameters $\Omega_{m0}$, $\Omega_{b0}$, $\Omega_{r0}$ and $H_0$ have been marginalized over.}
    \label{dblimits}
\end{table}
\end{center}

In most studies of dark energy models it is customary to ignore
curvature (e.g.\cite{DaveCaldwellSteinhardt, LiddleScherrer,
Dutta,Dutta1,Dutta2,Dutta3,Dutta4,Dutta5}), especially concerning
observational constraints. This practice is well motivated since
most inflationary scenarios predict a high degree of spatial
flatness, and furthermore the CMB data impose stringent
constraints on spatial flatness in the context of constant-$w$
models (for example a combination of WMAP+BAO+SNIa data
\cite{Komatsu:2008hk} provides the tight simultaneous constraints
$-0.0179\leq\Omega_{K0}\leq0.0081$ and $-0.12\leq1+w\leq0.14$,
both at 95\% confidence).

However, it is important to keep in mind that due to degeneracies
in the CMB power spectrum (see \cite{crooks} and references
therein), the limits on curvature depend on assumptions regarding
the underlying dark energy scenario. For instance, if instead of a
constant $w$  one assumes a linearly varying $w$ (that is
$w\(a\)=w_0+\(1-a\)w_a$), the error on $\Omega_{K0}$ is of the
order of a few percent, that is much larger
\cite{Wang2,Verde,Ichi1} (see \cite{Wright,Ichi2,Ichi3} for the
constraints on curvature for different parameterizations). The
authors of \cite{Ichi3} showed that for some models of dark energy
the constraint on the curvature is at the level of $5\%$ around a
flat universe, whereas for others the data are consistent with an
open universe with $\Omega_{K0}\sim0.2$. According to
\cite{Verde}, geometrical tests such as the combination of the
Hubble parameter $H(z)$ and the angular diameter distance
$D_A(z)$, using (future) data up to sufficiently high redshifts
$z\sim 2$, might be able to disentangle curvature from dark energy
evolution, though not in a model-independent way. Furthermore, in
\cite{Cortes,Virey} the authors highlighted the pitfalls arising
from ignoring curvature in studies of dynamical dark energy, and
recommended to treat $\Omega_{K0}$ as a free parameter to be
fitted along with the other model parameters. Lastly, note that in
the present work the spatial curvature plays a very crucial role,
since Ho\v{r}ava-Lifshitz cosmology coincides completely with
$\Lambda$CDM if one ignores curvature
\cite{Calcagni:2009ar,Kiritsis:2009sh}. Therefore, and following
the discussion above, we choose to treat $\Omega_{K0}$ as a free
parameter.
 \begin{figure}[htbp]
\begin{center}
\includegraphics[width=7cm]{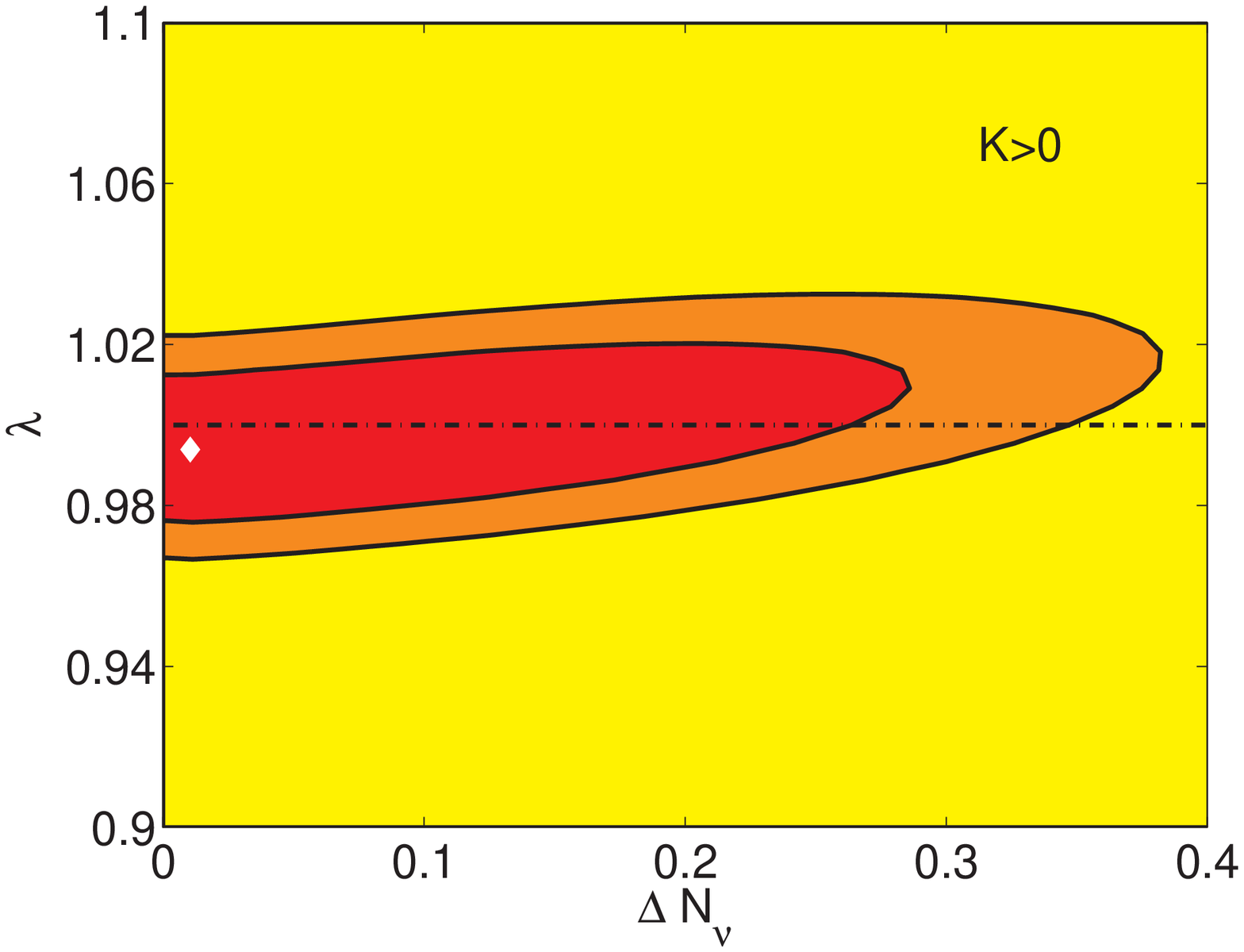}\\
 \includegraphics[width=7cm]{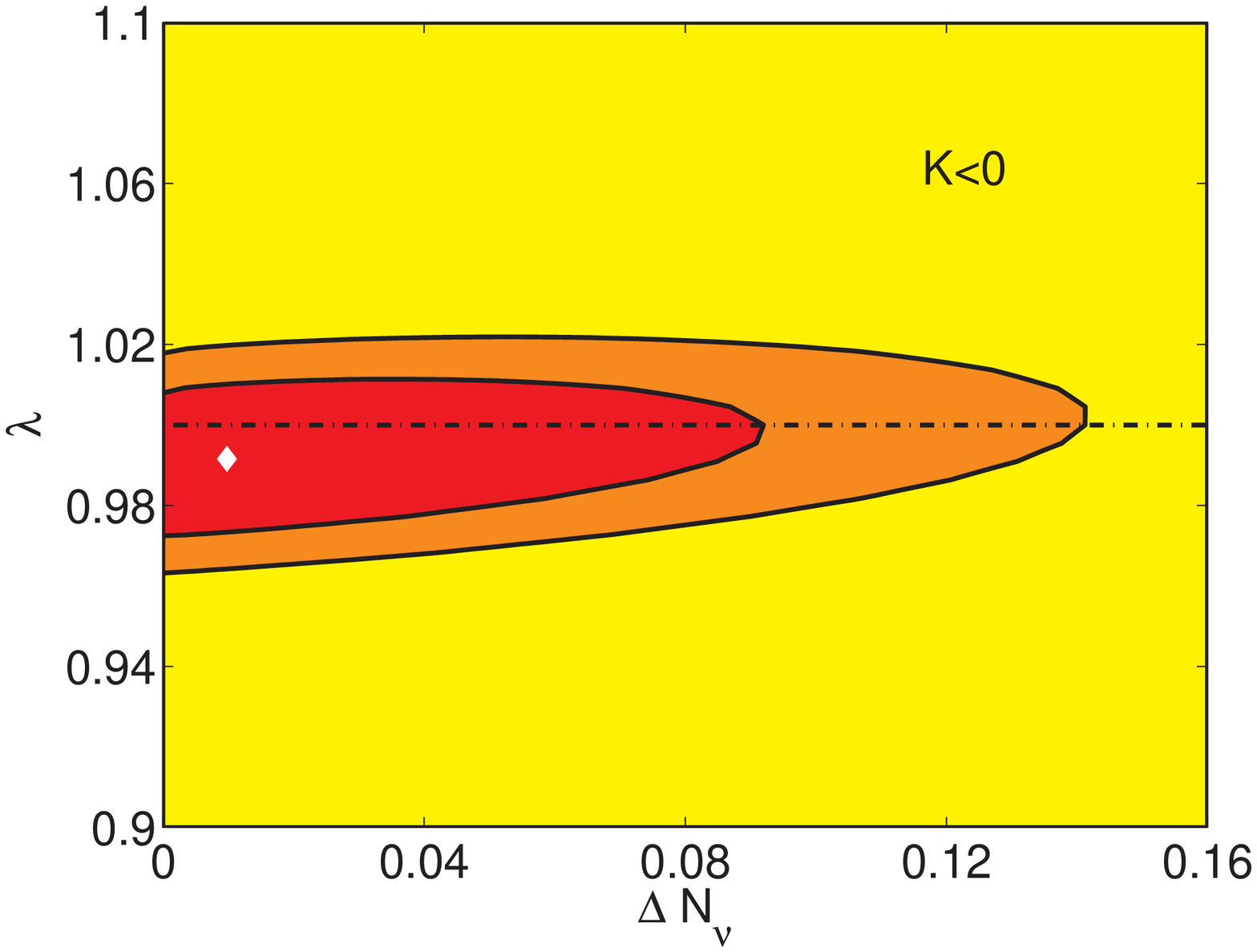}
\caption{ (Color online) {\it{Contour plots of $\lambda$ vs
        $\dn$ for negative
        ($K<0$) curvature, in the detailed-balance scenario,
        under SNIa, BAO and CMB observational data. The remaining
         parameters have been marginalized over (see text).
        The yellow (light) region
is excluded at the 2$\sigma$ level, and the orange (darker) region
is excluded at the 1$\sigma$ level.  The red (darkest) region is
not excluded at either confidence level. The white diamonds mark
the best-fit point, and the $\lambda=1$ line is drawn for
convenience. }}}
 \label{dbcontour}
 \end{center}
\end{figure}

In summary, the scenario at hand involves seven model parameters
namely the cosmological parameters $\Omega_{m0}$, $\Omega_{b0}$,
$\Omega_{k0}$, $\Omega_{r0}$ and $H_0$ and the model parameters
$\lambda$, $\omega$ and $\dn$, subject to constraint equations
(\ref{cond1}) and (\ref{cond2}). We marginalize over the
cosmological parameters $\Omega_{m0}$, $\Omega_{b0}$,
$\Omega_{r0}$ and $H_0$. Of the four remaining parameters, only
two are independent. We choose $\lambda$ and $\dn$  as our free
parameters. Once these are chosen, and for a given choice of
curvature, $\Omega_{K0}$ and $\omega$ are immediately fixed from
the constraint equations. In particular, $\omega$ can be
determined by eliminating $\Omega_{K0}$ from relations
(\ref{cond1}) and (\ref{cond2}):
\begin{align}
\label{omega}
\omega-&2\,{\rm sgn}\(\Omega_{K0}\)\sqrt{0.135\dn\,\Omega_{r0}\,\omega}\,+\,0.135\dn\Omega_{r0}\nonumber\\
+\,&2\[\frac{\Omega_{m0}+\Omega_{r0}}{3\lambda-1}\]-1=0.
\end{align}
This fractional-order equation in  $\omega$ can have one or two
roots. In the latter case we  use the larger one, since our goal is
to set an upper limit on $\omega$ (and hence on $\Lambda$).
$\Omega_{K0}$ can then be found from $\omega$ using (\ref{cond2}).

In Fig.~\ref{dbcontour} we display the likelihood-contours for the
free parameters $\lambda$ vs $\dn$ for both positive and negative
curvature. All other parameters have been marginalized over. The
details and the techniques of the fitting procedure can be found
in the Appendix of \cite{Dutta:2009jn}. Additionally,  in Table
\ref{dblimits} we summarize the $1\sigma$ limits on the parameter
values for the detailed-balance scenario. These bounds are in
agreement with the corresponding ones of our previous work
\cite{Dutta:2009jn}. Furthermore, as we observe, the data
constrain $\lambda$ to roughly $\lambda=1^{+0.01}_{-0.02}$ at the
$1\sigma$ level, while its best fit value is very close to $1$
($\lambda_{b.f}=0.006$) for both positive  and negative curvature.

\subsection{Constraints on Beyond-Detailed-Balance scenario}

\begin{center}
\begin{figure*}[ht]
\begin{tabular}{c@{\qquad}c}
\epsfig{file=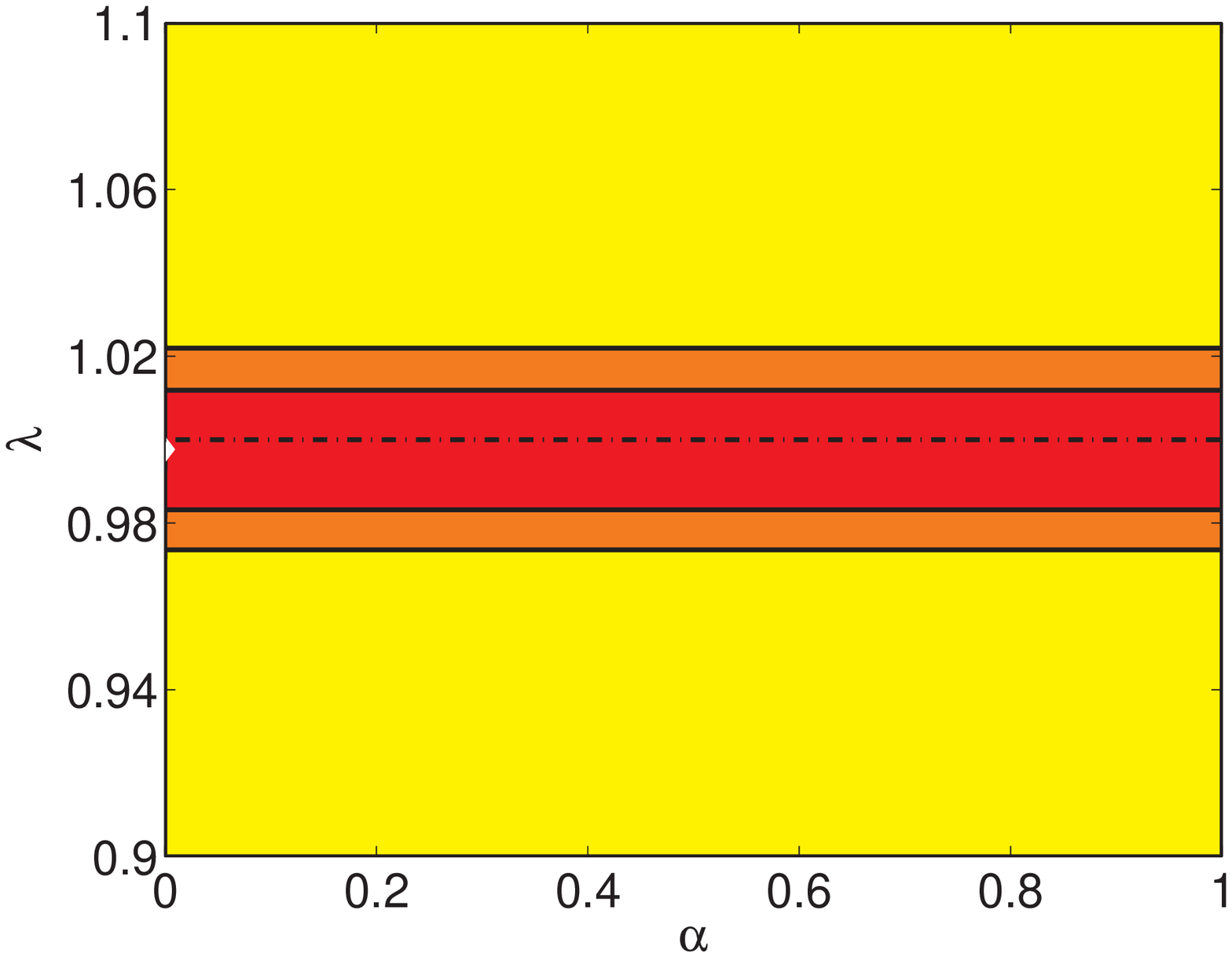,width=7 cm}&\epsfig{file=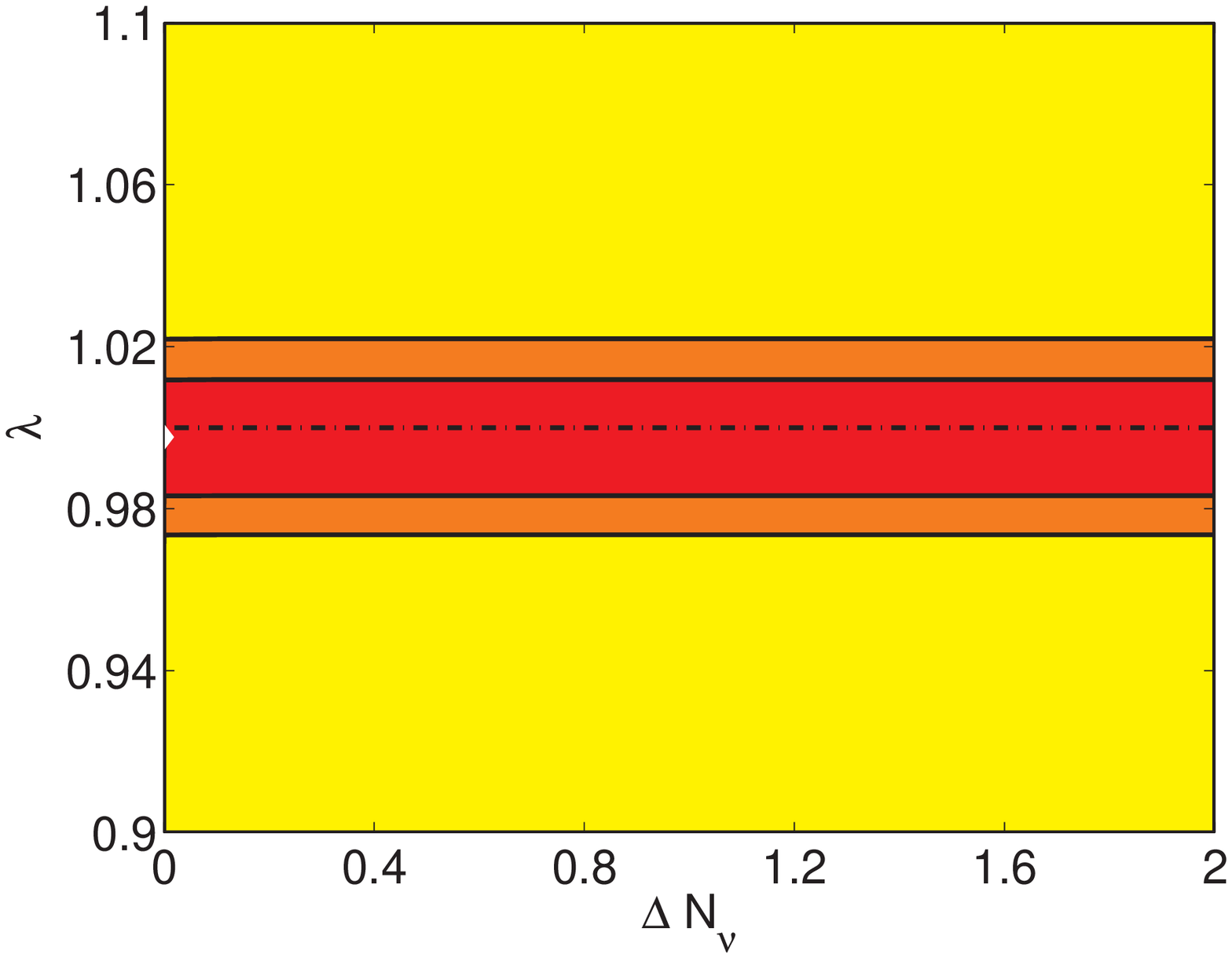,width=7 cm}\\
\epsfig{file=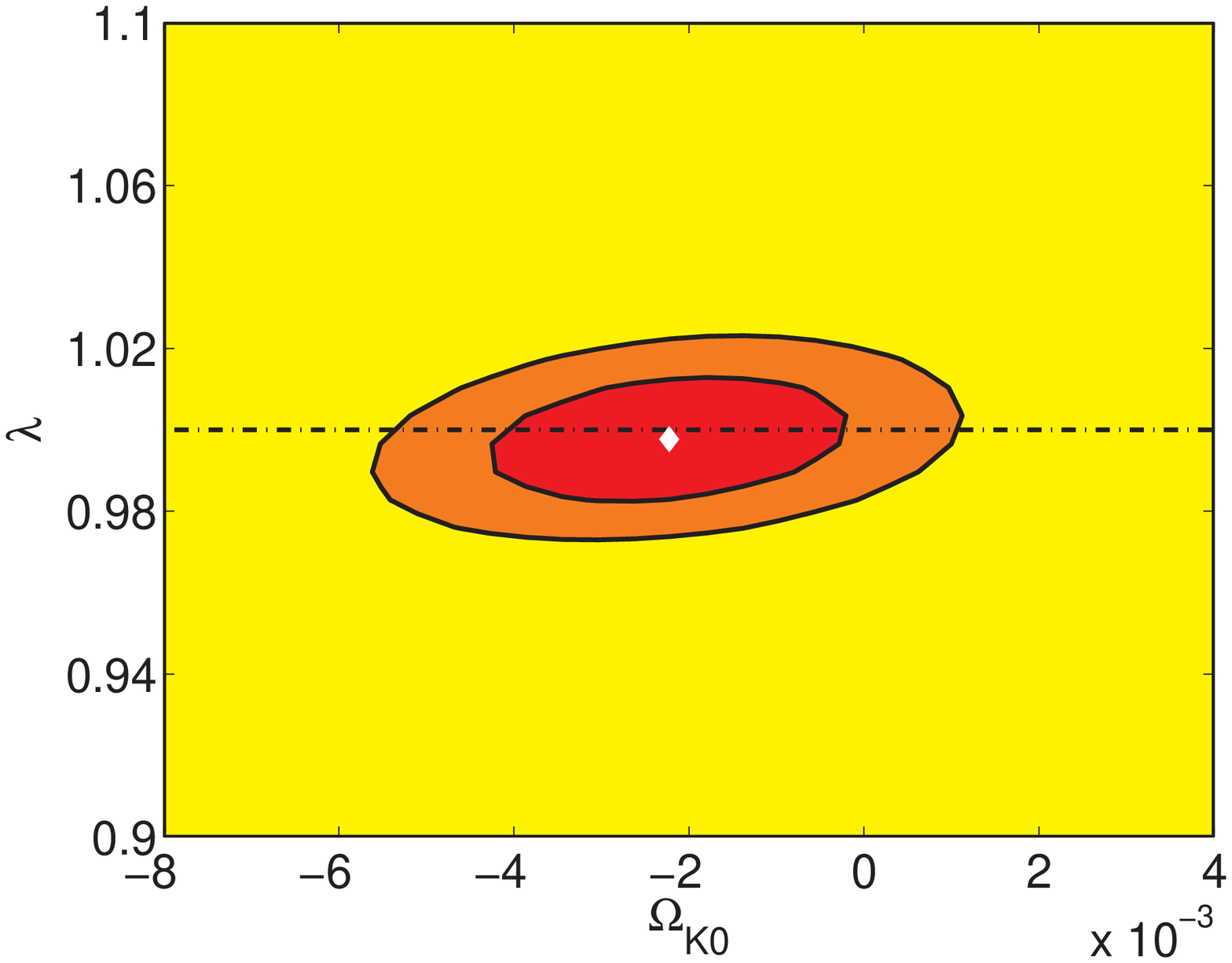,width=7
cm}&\epsfig{file=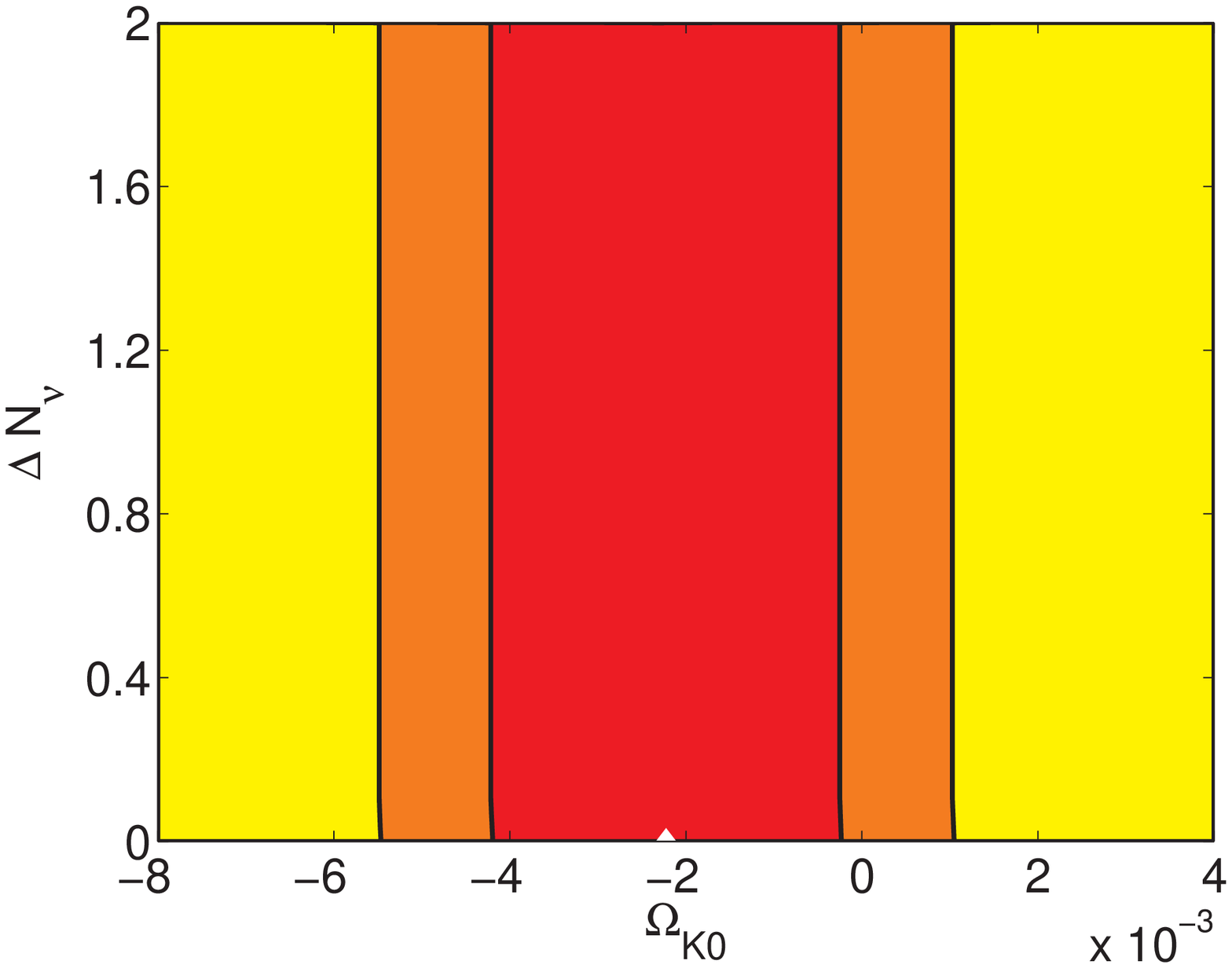,width=7 cm}
\end{tabular}
\caption{\label{bdbcontour}(Color online) {\it{Contour plots of
different pairs of free parameters in the beyond-detailed-balance
scenario,
        under SNIa, BAO and CMB observational data. In each case
         the parameters not included in the plots have been marginalized over. Color scheme as in Fig. \ref{dbcontour}.}}}
\end{figure*}
\end{center}

In units where $8\pi G_{\rm grav}=1$ relation (\ref{Ggravbdb}) gives
\begin{eqnarray}
\sigma_0=1/3\label{simpleconstantsndhelp}.
\end{eqnarray}
Using this value and following the procedure of the previous
subsection, the Friedmann equation (\ref{Fr1c}) can be written as
 \ba
   \label{Fr1c_a}
H^2&=&H_{0}^2\Big\{\frac{2}{(3\lambda-1)}\Big[\Omega_{m0}(1+z)^3+\Omega_{r0}(1+z)^4\Big]+\nonumber\\
&\ &+\Omega_{
K0}\(1+z\)^2+\nonumber\\
&\ &+\frac{2}{(3\lambda-1)}\Big[\omega_1+\omega_3
\(1+z\)^4+\omega_4 \(1+z\)^6\Big]\Big\},\ \ \ \ \ \
  \ea
where we have introduced the dimensionless
 parameters $\w_1$, $\w_3$ and $\w_4$, related to the model parameters $\sigma_1$, $\sigma_3$ and $\sigma_4$
 through:
\ba
\label{omegatosigma}
\omega_1&=&\frac{\sigma_1}{6H_0^2}\nonumber\\
\omega_3&=&\frac{\sigma_3 H_0^2\Omega_{ K0}^2}{6}\nonumber\\
\omega_4&=&-\frac{\sigma_4\Omega_{ K0}}{6}.
  \ea
   Additionally, we consider the
combination $\omega_4$ to be positive, in order to ensure that the
Hubble parameter is real for all redshifts. $\omega_4>0$ is
required also for the stability of the gravitational perturbations
of the theory \cite{Sotiriou:2009bx,Bogdanos:2009uj}. For
convenience we moreover assume $\sigma_3\geq0$, that is
$\omega_3\geq0$.

In summary, the present scenario involves the following
parameters: the cosmological parameters $H_0$, $\Omega_{m0}$,
 $\Omega_{K0}$, $\Omega_{b0}$, $\Omega_{r0}$, and the model parameters $\lambda$, $\w_1$, $\w_3$
and $\w_4$. Similarly to the detailed-balance section these
are subject to two constraints. The first one arises from the
Friedman equation at $z=0$, which leads to
  \be
     \label{ndbcond1}
\frac{2}{(3\lambda-1)}\Big[\Omega_{m0}+\Omega_{r0}+\w_1+\w_3+\w_4\Big]+\Omega_{K0}=1.
  \ee
  This constraint eliminates the parameter $w_1$.
The second constraint arises from BBN considerations. The term
involving $\w_3$ represents the usual dark-radiation component. In
addition, the $\w_4$-term represents a kination-like component (a
quintessence field dominated by kinetic energy
\cite{kination,kination1}). If $\dn$ represents the BBN upper
limit on the total energy density of the universe beyond standard
model constituents, then as we show in the Appendix of
\cite{Dutta:2009jn} we acquire the following constraint at the
time of BBN ($z=z_{\rm BBN}$)
\cite{BBNrefs,BBNrefs1,BBNrefs2,Malaney:1993ah}:
 \be
\label{ndbcond2} \w_3+\w_4\(1+z_{\rm
BBN}\)^2=\w_{3\text{max}}\equiv0.135\dn\Omega_{r0}. \ee
It is clear that BBN imposes an extremely strong constraint on
$\w_4$, since its largest possible value (corresponding to
$\w_3=0$) is $\sim 10^{-24}$. Finally, $\w_{3\text{max}}$ denotes
the upper limit on $\w_3$. In the following, we use expression
(\ref{ndbcond2}) to eliminate $\w_4$. For convenience, instead of
$\w_3$ we define a new parameter
\begin{equation}
\alpha\equiv  \frac{\w_3}{\w_{3\text{max}}},
  \end{equation}
 which has the interesting physical meaning of denoting
the ratio of the energy density of the Ho\v{r}ava-Lifshitz dark
radiation  to the total energy density of Ho\v{r}ava-Lifshitz dark
radiation and kination-like components at the time of BBN.

We use relation (\ref{ndbcond2}) to eliminate $\w_4$ in favor of
$\alpha$ and $\dn$, and treat  $\lambda$,  $\alpha$, $\Omega_{K0}$
and $\dn$ as our free parameters, marginalizing over $H_0$,
$\Omega_{m0}$, $\Omega_{b0}$ and $\Omega_{r0}$. Using the combined
SNIa+CMB+BAO data, we construct likelihood contours for different
combinations of the above  parameters, which are presented in Fig.
\ref{bdbcontour}. In each case, the free parameters not included
in the plot have been marginalized over. The details and the
techniques of the construction have been described in the Appendix
of \cite{Dutta:2009jn}, with the only difference being that in the
present work WMAP 7-year data \cite{Komatsu:2010fb}   and the more
recent Constitution Supernovae dataset \cite{hicken} have been
used. In Table \ref{ndblimits} we summarize the $1\sigma$ limits
on the parameter values for the beyond-detailed-balance scenario.
\begin{center}
\begin{table}[ht]
    \centering
    \scalebox{.92}{
        \begin{tabular}{|c|c|c|c|}
        \hline
        \textbf{$\Omega_{K0}$} &    \textbf{$\dn$} &  \textbf{$\alpha$}       & \textbf{$\lambda$}  \\\hline
        $(-0.01,0.01)$         & $(0,2)$         & $(0,\, 1)$            &  $(0.98,\,1.01)$    \\\hline
        \end{tabular}}
    \caption {1$\sigma$ limits on the free parameters of the beyond-detailed-balance scenario. The cosmological parameters $\Omega_{m0}$, $\Omega_{b0}$, $\Omega_{r0}$ and $H_0$ have been marginalized over.}
    \label{ndblimits}
\end{table}
\end{center}

Furthermore, as we observe, in $1\sigma$ confidence the running
parameter $\lambda$ of Ho\v{r}ava-Lifshitz gravity is restricted
to the interval $|\lambda-1|\lesssim0.02$, for the entire allowed
range of $\omega_3$ (that is of $\sigma_3$). Finally, the best fit
value for $\lambda$ restricts $|\lambda-1|$ to much more smaller
values, namely $|\lambda_{b.f}-1|\approx0.002$.

\section{Conclusions}
\label{conclusions}

In this work we have constrained the running parameter $\lambda$
of Ho\v{r}ava-Lifshitz gravity using observational SNIa, BAO and
CMB  data as well as considerations from BBN. In order to be
general enough we have not used any theoretical argument to a
priori restrict $\lambda$  in any specific interval, handling it
as a completely free parameter. Additionally, we have performed
our investigation under the detailed-balance condition, as well as
under its relaxation. Finally,  we have included the matter and
radiation sectors following the usual effective fluid approach. We
stress that we have let all the parameters of the theory to vary,
performing an overall and holistic investigation, and we have not
followed the naive approach, that is to keep everything fixed and
vary only $\lambda$ in order to match observations.

As we showed, Ho\v{r}ava-Lifshitz cosmology, either with or
without the detailed-balance condition, can be compatible with
observations. We constructed the likelihood-contours for the
involved free parameters, and we found that in $1\sigma$ level
$\lambda$ is restricted to $|\lambda-1|\lesssim0.02$. As expected,
these bounds are one order of magnitude tighter than the
corresponding ones arising from the consideration of primordial
He$^4$-abundance measurements in order to extract a crude upper
bound \cite{Blas:2009qj,Papazoglou:2009fj}.  Finally, concerning
the best fit value for $\lambda$ we obtain
$|\lambda_{b.f}-1|\approx0.006$ for the detailed balance and
$|\lambda_{b.f}-1|\approx0.002$ for the beyond-detailed-balance
cases.  The aforementioned features act as additional arguments in
favor of the present holistic investigation, where all the
parameters of the theory are allowed to vary, comparing to the
naive or partial ones in which only $\lambda$ varies. Lastly, it
is interesting to note that these results, arising from
cosmological observations, are relatively close to the preliminary
parametrized post-Newtonian (PPN) estimations for
Ho\v{r}ava-Lifshitz gravity ($0<|\lambda-1|\lesssim4\times
10^{-7}$ \cite{Blas:2009ck}), despite the fact that the Solar
System measurements (that lie at the basis of PPN parameters
\cite{PPN,PPN1,PPN2,PPN3}) are much more accurate than the
cosmological ones.

In summary, as expected, we found that the value of the running
parameter $\lambda$ of Ho\v{r}ava-Lifshitz gravity is restricted
to a very tight window around its IR value.

Finally, we should mention that although the present analysis
provides the bounds for the running parameter, it does not
enlighten the discussion about the possible conceptual problems
and instabilities of Ho\v{r}ava-Lifshitz gravity,  nor it can
address the questions concerning the validity of its theoretical
background, which is the subject of interest of other studies. The
present work just faces the problem from the phenomenological
point of view, which is necessary but not sufficient, and thus its
results can been taken into account only if Ho\v{r}ava-Lifshitz
gravity passes successfully the aforementioned theoretical tests.

\section{Acknowledgements}
The authors wish to thank Y. F. Cai for useful discussions.

\end{document}